\documentclass[usenatbib,usegraphicx,a4paper,useAMS]{mn2e}
\usepackage{amssymb,url,times}
\usepackage{graphicx,epsfig,natbib}
\usepackage{txfonts}
\usepackage{epstopdf}
\usepackage{color}

\def\be{\begin{equation}}
\def\ee{\end{equation}}

\title[Spin alignment during black hole mergers]{Black hole mergers: do gas discs lead to spin alignment?}

\author[Lodato \& Gerosa]{Giuseppe Lodato$^1$ and Davide Gerosa$^{1,2}$ \\
$^1$Dipartimento di Fisica, Universit\`a Degli Studi di Milano, Via Celoria, 16, Milano, 20133, Italy\\
$^2$Department of Physics and Astronomy, The University of
Mississippi, University, MS 38677, USA
}
\date{Submitted: Revised:  Accepted:}
\pagerange{\pageref{firstpage}--\pageref{lastpage}} \pubyear{2010}

\begin{document}
\label{firstpage}
\bibliographystyle{mn2e}
\maketitle

\begin{abstract}
In this Letter we revisit arguments suggesting that the Bardeen--Petterson
effect can coalign the spins of a central supermassive black hole binary
accreting from a circumbinary  (or circumnuclear) gas disc. We improve on previous estimates by
adding the dependence on system parameters, and noting that the nonlinear
nature of warp propagation in a thin viscous disc affects alignment. This
reduces the disc's ability to communicate the warp, and can severely reduce
the effectiveness of disc--assisted spin alignment. We test our predictions with a Monte Carlo 
realization of random misalignments and accretion rates and we find that the outcome 
depends strongly on the spin magnitude. We estimate a generous upper
limit to the probability of alignment by making assumptions which favour it
throughout. Even with these assumptions, about 40\% of black holes with $ a
\gtrsim 0.5$  do not have time to align with the disc. If the residual misalignment is not small and it is 
maintained down to the final coalescence phase this can give a powerful recoil velocity to the 
merged hole. Highly spinning black holes are thus more likely of being subject to strong recoils,
the occurrence of which is currently debated.
\end{abstract}

\begin{keywords}{accretion, accretion discs -   
          black hole physics - 
          galaxies: evolution -
          galaxies:  nuclei }
          \end{keywords}

\section{Introduction}

The cosmic evolution of the spin of supermassive black holes (SMBH)
residing in galaxy centers has several important astrophysical
implications. By regulating the efficiency of conversion of matter
into radiation, $\epsilon$, the spin magnitude $a$ effectively
determines the ability of black holes to grow rapidly by
accretion. Defining the Eddington time as
\begin{equation}
t_{\rm Edd} = \frac{\kappa c}{4\pi G}\approx 4.5\times 10^8\mbox{yrs},
\end{equation}
where $\kappa$ is the Thomson opacity, $c$ the speed of light and $G$ the
gravitational constant, a black hole accreting continuously at the Eddington
rate grows in mass exponentially on a timescale $\tau = \epsilon t_{\rm
  Edd}/(1-\epsilon)$. Rapidly spinning black holes have $\epsilon\approx 0.5$,
and thus $\tau\approx t_{\rm Edd}$, while for non-rotating black holes
$\tau\approx 0.06 t_{\rm Edd}$. If most holes spin rapidly, the existence of
luminous quasars at redshift $\gtrsim 7$, when the age of the Universe was $\sim
2t_{\rm Edd}$, \citep{fan04,fan06,mortlock11} requires very massive ($\gtrsim
10^5{\rm M_{\odot}}$) seed black holes \citep[e.g.][]{VR2005}, that might form through direct collapse of primordial gas \citep{LN06}. Alternatively,  \cite{king06,king07} and \cite{kph08} argued instead that the spins (and so
$\epsilon$) could remain relatively low, if most SMBH accrete through a
sequence of randomly oriented accretion events, in what is known as the
chaotic accretion picture. 


Black hole spin is also important in a completely different context.
The relative orientation of spins during binary black hole coalescence
determines the waveform of the gravitational radiation that may be
detected by planned gravitational wave observatories such as
NGO-eLISA. Moreover, when the spins of the two merging black holes are
significantly misaligned to each other, the remnant black hole can
receive recoil velocities up to $\sim 4000$ km/sec
\citep{campanelli2007a,campanelli2007b,lousto11}. There have been
claims of the detection of several fast recoiling black holes via their
emission lines \citep{komossa08,civano12}, but the interpretation is
unclear \citep{bogdanovic09,dotti09,blecha12} (see \citealt{komossa12}
for a review). The recoil can also lead to a significant brightening
of the accretion disc
\citep{schnittman08,rossi10,corrales10,anderson10} to
near--Eddington luminosities, either in the infrared
\citep{schnittman08} or at higher frequencies (UV to X-rays) if the
radiation is not thermalized \citep{rossi10}.

Black hole mergers are expected to occur in gas rich
environments, and circumbinary gas discs may play a key role in
bringing the binary to separations close enough for gravitational wave
emission to drive the final coalescence. The effectiveness of such
gas--assisted mergers is debated. Galaxy--scale simulations appear to
imply that in the presence of gas a black hole pair might reach 
separations of the order of 0.01 pc and form a binary
within $1-5 \times 10^7$ years \citep{escala05,dotti09b}. However,
higher--resolution computations \citep{lodato09} show that, at
distances of the order of 0.01 pc, the circumbinary disc probably
becomes self--gravitating and fragments, severely limiting the
possibility of mergers within a Hubble time. In contrast, in the
context of the chaotic accretion picture of \citet{king06},
\citet{nixon11a} have shown that a sequence of accretion episodes
where the circumbinary disc can be either co-- or counter--aligned
with the binary is much more effective than a simple prograde disc in
bringing the binary to coalescence.

Gas discs can also affect the mutual orientation of the spins of the two black
holes through the Bardeen--Petterson (BP) effect \citep{bardeen75,pappringle83}. 
Lense--Thirring precession and viscous
torques in a disc around a black hole tend to co-- or counteralign its spin
with the disc axis \citep{LP06,klop05}.  In a binary black hole system, each hole may have an
individual disc around it, whose initial orientation agrees with that of the
circumbinary disc. Then the BP effect can co-- or counteralign the two spins.
Clearly, for this to work, the orientation of the circumbinary disc, which is
feeding the individual black hole discs, must stay roughly constant. (This
does not hold in the chaotic accretion picture, cf
\citealt{nixon11a,nixon11b,n12}.)

\citet{bogdanovic07} (hereafter BRM) make order--of--magnitude estimates of
the BP effect. They consider the alignment timescale for a single black hole
and its accretion disc. BRM find that this is very short, and conclude that
each black hole is effectively aligned with its disc and so that the two black
hole spins are each co-- or counter--aligned with the common circumnuclear 
disc at the time when they form a binary. \cite{perego09} 
 improve their study finding slightly longer timescale. The
numerical simulations of \citet{dotti10} appear to support this picture,
although with some differences, depending on whether the circumnuclear disc is
warm or cold, with the coldest (and thinnest) disc providing more effective
alignment.

The estimates of BRM have a number of shortcomings. First, as
mentioned above, their argument assumes that the discs surrounding
each black hole share their orientations with the circumbinary disc,
and that this does not change with time. If the accretion process
during the merger is chaotic, both assumptions may be invalid. But
even within the framework where the disc orientations track the binary
orbit, the alignment timescale may not be as short as implied by
BRM. Indeed, 
BRM evaluate the accretion timescale is by considering the flow properties at the Bondi radius, at a
distance of 40~pc from the hole. However, the viscous time at
these distances from the hole is much larger than a Hubble time for a thin disc, and still larger than 
the binary evolution timescale if the disc is assumed to be thick. Thus, such estimates may not accurately describe the flow properties at the warp radius, located at a few hundreds Schwarzschild radii. Secondly,
and most importantly, both BRM and \cite{perego09} assume that the warp diffusion
coefficient, usually called $\nu_2$, is given by
$\nu_2/\nu_1=1/2\alpha^2$, where $\nu_1$ is the disc viscosity and
$\alpha$ the standard Shakura-Sunyaev parameter
\citep{shakura73}. This is correct only to first order for small
amplitude warps \citep{pappringle83}, while large amplitude warps
diffuse much less efficiently \citep{ogilvie99,LP10} (note also that
the simulations by \citealt{dotti10} assume that the warp diffusion
coefficient is independent of amplitude). It is thus possible that
large initial misalignments stay misaligned throughout the merger.

In this Letter we revisit the arguments of BRM and calculate a lower
limit to the probability that the black hole spin remains misaligned
(i.e. it does not have sufficient time to reach the aligned configuration)
with its own disc at the point when gravitational wave reaction begins
to control the merger process. In such a case it is very unlikely that
the the two spins in a binary will end up aligned  unless the initial misalignment is very small.
 We repeat that even
in cases where we predict local alignment, a mutual coalignment of the
two spins is only possible if the merger does not occur through a
sequence of chaotic accretion events.

The paper is organized as follows: in section \ref{sec:model} we
describe the basic assumptions of our model and the procedure we adopt
to compute the alignment probability. In section \ref{sec:results} we
discuss our main results for a suitable choice of parameters, and in
section \ref{sec:conclusions} we draw our conclusions.

\section{Modeling spin alignment}
\label{sec:model}

\citet{natarajan98} estimate the alignment timescale for the BP effect
in the linear case where $\nu_2/\nu_1=1/2\alpha^2$. BRM use these
results to estimate spin alignment. In general, one can define an
$\alpha_2$ coefficient corresponding to the diffusion coefficient
$\nu_2$, such that $\nu_2/\nu_1=\alpha_2/\alpha$. The value of
$\alpha_2$ can be computed based on the nonlinear theory of warp
propagation of \citet{ogilvie99}. This uses conservation laws to work
out the connections between the viscosity coefficients, assuming that
the local viscosity is isotropic. The validity of this theory has been
demonstrated numerically by \citet{LP10}. For small warps, we have:
\begin{equation}
\frac{\alpha_2}{\alpha}=\frac{1}{2\alpha^2}\frac{4(1+7\alpha^2)}{4+\alpha^2}.
\label{eq:prediction2}
\end{equation}
The above equation is, however, inadequate for large amplitude
warps. \citet{ogilvie99} provides a slowly converging Taylor series
including additional amplitude dependent terms, but the full
nonlinear result must be computed numerically.

The warp amplitude is measured by the parameter \mbox{$\psi =
  \mbox{d}\beta/\mbox{d}\ln R$}, where $\beta$ is the local inclination of the
disc. To compute the warp evolution accurately, one would thus not only need
to know the misalignment between the outer disc axis and the black hole spin,
$\theta$, but also how steep its gradient is. This clearly requires a detailed
calculation of the shape of the disc. For our preliminary assessment of the
effect, we simply assume that the warp is a smooth function, varying over a
distance of the order of the disc size, and so approximately $\psi\approx
 \theta$. This is a conservative assumption, since a
steeper warp profile would increase the warp amplitude and thus reduce further
the effectiveness of warp propagation. Indeed for sufficiently high
inclination or low viscosity the disc may break, i.e. make a sharp transition
between two distinct planes \citep{NK12}.

The alignment timescale for small warps and a steady state disc is given by
\citep[e.g.][]{sf96,natarajan98,LP06} as
\begin{equation}
t_{\rm align}=
3a\frac{\nu_1}{\nu_2}\frac{M}{\dot{M}}\left(\frac{R_{\rm S}}{R_{\rm
    w}}\right)^{1/2},
\label{eq:align}
\end{equation}
where $M$ is the black hole mass, $a$ the hole spin parameter, $\dot{M}$ the
accretion rate through the disc, $R_{\rm S}=2GM/c^2$ is the Schwarzschild
radius, and $R_{\rm w}$ is the warp radius. 
In (\ref{eq:align}) the
warp radius $R_{\rm w}$ is taken as the point at which the warp diffusion
timescale equals the Lense--Thirring timescale, so that
\begin{equation}
R_{\rm w} = \frac{2GJ_{\rm h}}{c^2\nu_2},
\end{equation}
where $J_{\rm h}=aGM^2/c$ is the black hole angular momentum. The ratio $R_{\rm
  w}/R_{\rm S}$ is given by (cf. the analogous expression in
\citealt{natarajan98}):
\begin{equation}
\frac{R_{\rm w}}{R_{\rm S}}=\frac{1}{2^{1/3}}
\left(\frac{a}{\alpha_2}\right)^{2/3}\left(\frac{H}{R}\right)^{-4/3},
\end{equation}
where $H/R$ is the aspect ratio of the disc. Inserting this in
equation \ref{eq:align} and scaling the accretion rate in units of the
Eddington value $\dot{M}_{\rm Edd}=\epsilon M/t_{\rm Edd}$ we obtain:
\begin{eqnarray}
\label{eq:timescale}
&\displaystyle t_{\rm align}\approx 3.37 \alpha
  \left(\frac{a}{\alpha_2}\right)^{2/3}\left(\frac{\dot{M}}{\dot{M}_{\rm
      Edd}}\right)^{-1}\left(\frac{H}{R}\right)^{2/3}\epsilon t_{\rm
    Edd}\approx\\ \nonumber &7\times 10^6
\displaystyle  \left(\frac{a}{\alpha_2}\right)^{2/3}
\left(\frac{\alpha}{0.1}\right)\left(\frac{\dot{M}}{0.1\dot{M}_{\rm
      Edd}}\right)^{-1}
\left(\frac{H/R}{0.01}\right)^{2/3}\left(\frac{\epsilon}{0.1}\right)\,{\rm
  yr},
\end{eqnarray}
where the various parameters have been scaled to typical values. Note
that this timescale is only a factor 2 smaller than the typical
estimates for the shrinking time in a gas-rich environment, which is $\sim 10^7$ years
\citep{escala05,dotti09b}. For small warps, $\alpha_2$ can become very
large, so the alignment timescale is much smaller than the
shrinking time. The same also occurs for slowly spinning black holes,
for which both $a$ and the accretion efficiency $\epsilon$ are
small. Conversely, for maximally spinning black holes,
$a\approx 1$, for which $\epsilon\approx 0.4$, 
alignment for large warps might require a timescale longer than the
shrinking time.

\begin{figure}
  \centerline{\epsfig{figure=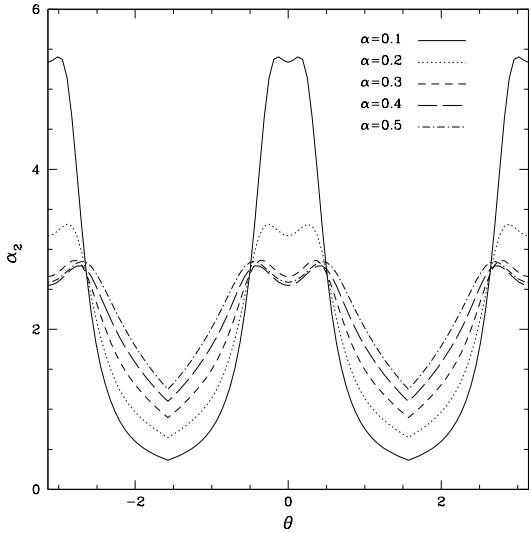,width=0.43\textwidth}}
\caption{The value of the diffusion coefficient $\alpha_2$ as a
  function of the misalignment $\theta$ for several choices of
  $\alpha$=0.1 (solid line), 0.2 (dotted line), 0.3 (dashed line), 0.4
  (long-dashed line) and 0.5 (dot-dashed line). While for small
  misalignments ($\theta\approx 0$), $\alpha_2$ can be quite large,
  its value drops significantly (even below unity) when the
  misalignment approaches $\pi/2$.}
\label{fig:alpha2}
\end{figure}

\begin{figure*}
\begin{center}
\includegraphics[width=0.85\columnwidth]{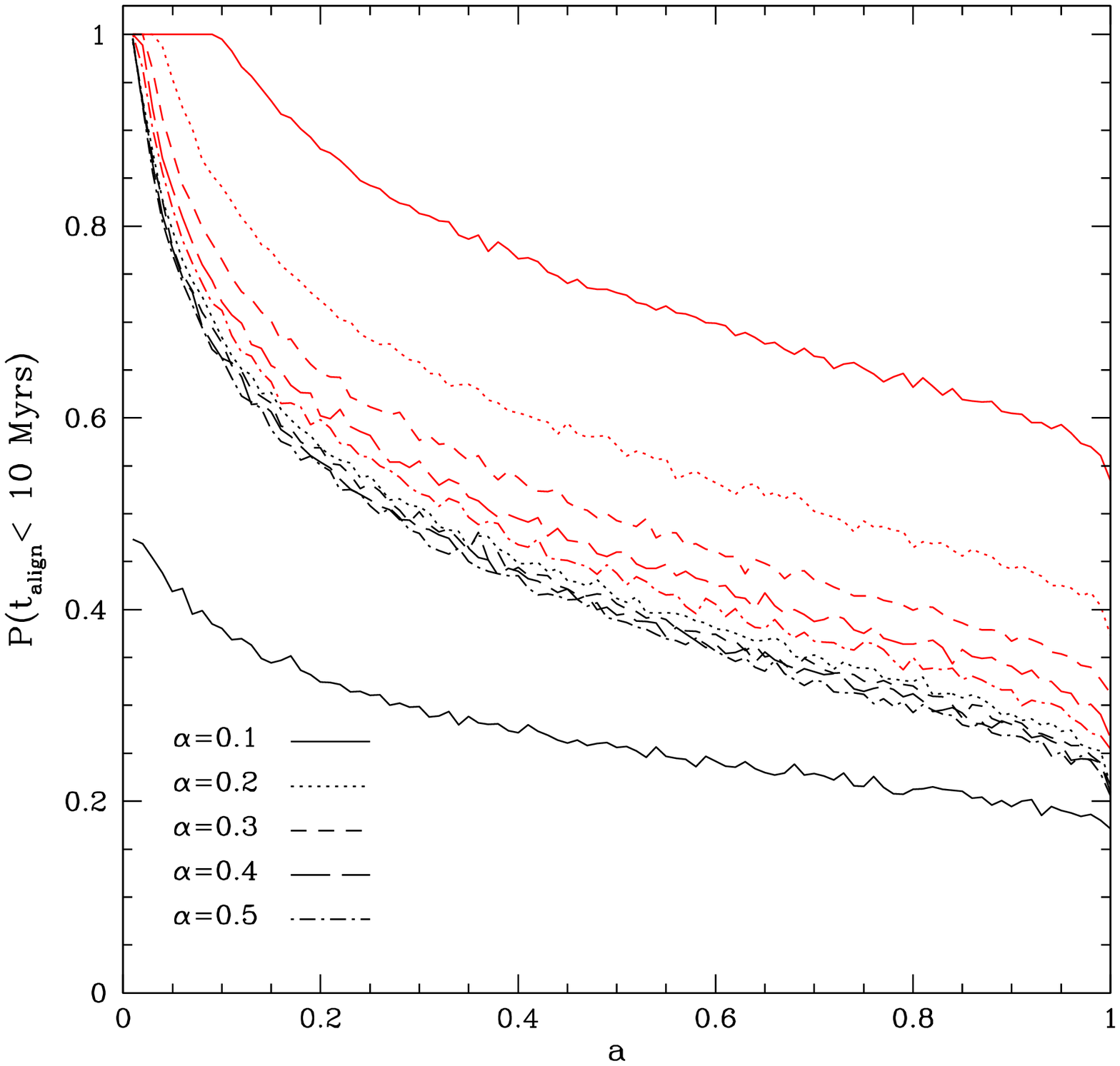}
\includegraphics[width=0.85\columnwidth]{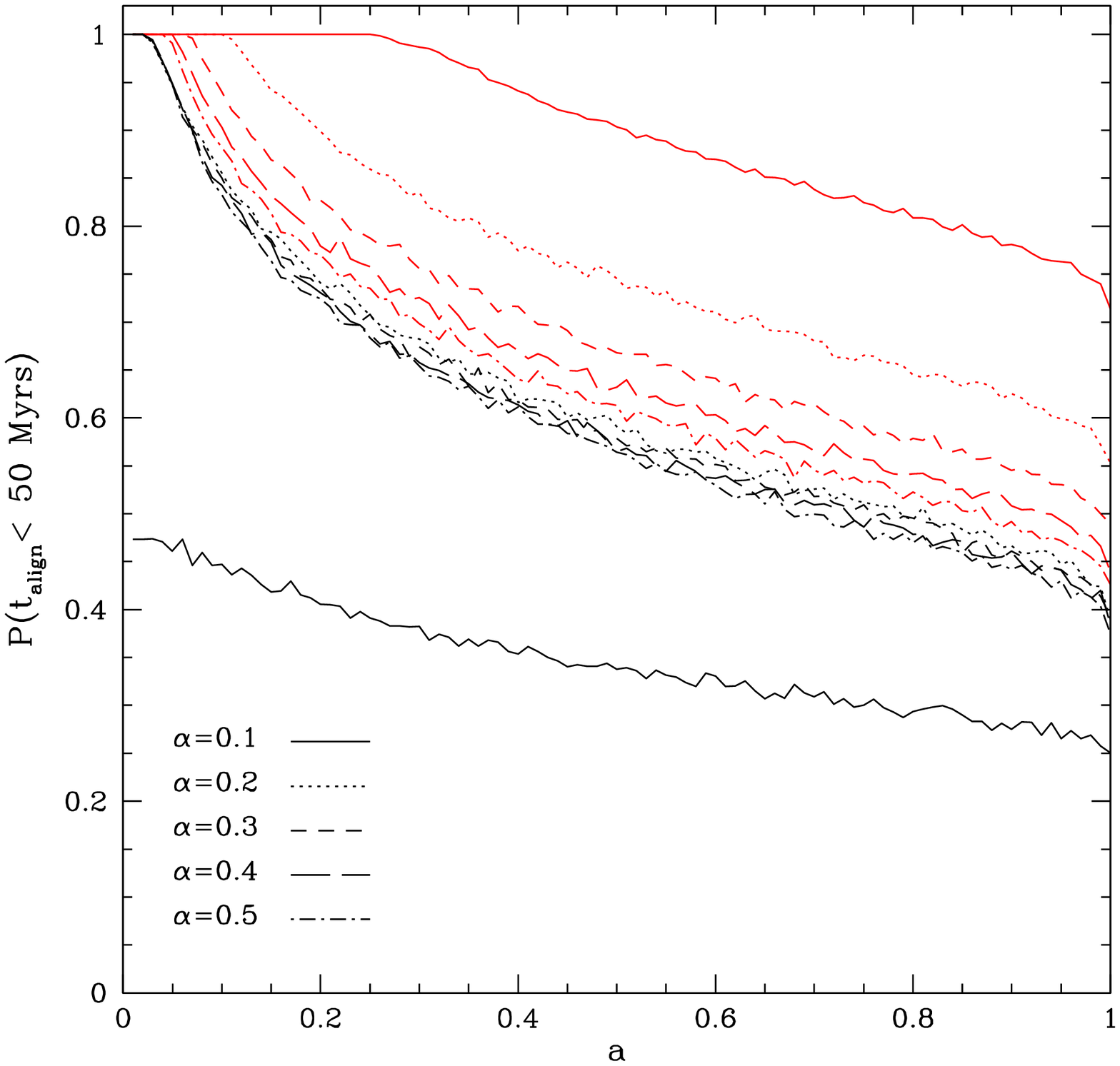}
\caption{Probability that the alignment time $t_{\rm align}$ is
  smaller than 10 Myrs (left panel) or 50 Myrs (right panel), as a
  function of the spin parameter $a$, for various choices of $\alpha$,
  using the same notation as in Fig. \ref{fig:alpha2}. For large
  values of $a$ a sizeable fraction of systems are likely to have
  residual spin misalignments  after a binary shrinking time. The red lines show the corresponding probabilities in the case where we use Eq. (2) to compute $\alpha_2$ and we thus do not consider the effect of the non-linearity of the warp on the diffusion coefficients. In this case, there is a much stronger dependence on $\alpha$ and for small $\alpha$ alignment is much more efficient.}
  \end{center}
\label{fig:probs}
\end{figure*}

\citet{natarajan98} also evaluate $H/R$ as a function of the main disc
parameters (black hole mass and accretion rate). However, the
dependence turns out to be rather weak, and so we have just chosen to
keep that number fixed at a representative value of 0.01. In (\ref{eq:timescale}) we scaled
the Eddington ratio $\dot{M}=0.1\dot{M}_{\rm Edd}$, which is typical for accreting black holes in AGN. Also this assumption is favourable for the alignment process since it is expected that BH in binary systems accrete ad a reduced rate due to binary torques (typically at 10 per cent of the unperturbed rate, \citealt{arty96}) and because fragmentation in the outer disc might lead to significant star formation \citep{lodato09}. However, in our calculation we explore a wider range of $\dot{M}$. 
 
In our calculation
we take into account the relationship between the spin parameter $a$
and the accretion efficiency $\epsilon(a)$, based on the formalism of
\citet{bardeen73} (cf. \citealt{kph08}, their fig. 5):
\begin{equation}
\epsilon= 1- \frac{r_{\mathrm{isco}}^{3/2}  -  2  r_{\mathrm{isco}}^{1/2}  \pm a} {r_{\mathrm{isco}}^{3/4} \left( r_{\mathrm{isco}}^{3/2}  - 3  r_{\mathrm{isco}}^{1/2}  \pm 2 a \right)^{1/2}}  ,
\end{equation}
where $r_{\mathrm{isco}}=R_{\rm isco}c^2/GM$ is the radius of the innermost stable orbit in units of the gravitational radius and is given by
\begin{equation}
R_{\mathrm{isco}}= \frac{GM}{c^2} \left[3 + Z_2 \mp  \left( 3-Z_1 \right)^{1/2} \left( 3+Z_1 +2 Z_2 \right)^{1/2}    \right] ,
\end{equation}
\begin{equation}
Z_1= 1+ \left( 1 - a^2 \right)^{1/3} 
\left[  \left(1+ a\right)^{1/3}   + \left(1- a \right)^{1/3} \right] \;,
\end{equation}
\begin{equation}
Z_2= \left(3 a^2 + Z_1^2\right)^{1/2} \; ,
\end{equation}
and the upper/lower sign refers to particle orbiting in prograde/retrograde orbits with respect to the rotation of the hole.

Finally, for any given misalignment $\theta$, we compute
$\alpha_2(\alpha,\theta)$ numerically, using the full nonlinear
theory of \citet{ogilvie99}. We plot in Fig. \ref{fig:alpha2} the
resulting value of $\alpha_2$ as a function of $\theta$ for various
choices of $\alpha$, which shows that while for small misalignments
($\theta\approx 0$), $\alpha_2$ can be quite large, its value drops
significantly (even below unity) when the misalignment approaches
$\pi/2$.

\section{Results}
\label{sec:results}

We compute the probability of alignment through a Monte Carlo realization of
$N=10^4$ events. The initial angles $\theta$  are randomly distributed  between $-\pi$ and $\pi$,
thus including also cases where the disc and the black hole end up in a
counteraligned configuration. We generate the angles  uniformly in $\cos \theta$ to have an isotropic distribution in three dimensions. For each event we generate a random value of the accretion rate $\dot{M} / \dot{M}_{\rm Edd}$ between $10^{-4}$ and $1$ (the same range as explored by \citealt{perego09}), with uniform logarithmic distribution.
For each choice of the remaining two free
parameters $\alpha$ and $a$, we thus obtain the probability distribution of
$t_{\rm align}$: we compute this distribution varying over 5 different values
of $\alpha$ and 100 different values of $a$. Note that, for large $\theta$ and
small $\alpha$, the theory of \citet{ogilvie99} would fomally predict a
negative azimuthal viscosity coefficient. 
The behaviour of the disc in such cases is unclear, but
since the viscosity must already have passed through small positive values it
is likely that it has in reality broken into two distinct planes, thus making
alignment even slower \citep{LP10,NK12}. When this occurs, we simply assume
that the real alignment time becomes much larger than any other timescale in
the problem.

We plot in Fig. \ref{fig:probs} with black lines the probability that the alignment time is
smaller than 10 Myrs (left panel) and 50 Myrs (right panel) as a function of
the spin parameter $a$, for various choices of $\alpha$ (with the same
notation as Fig. \ref{fig:alpha2}). The values of 10 and 50 Myrs are taken as
proxies for the shrinking timescale of the binary \citep{escala05,dotti09b}.
The plots demonstrate the general features already discussed in section
\ref{sec:model}. For small spin parameters alignment is indeed efficient and
we expect most spins to align or counteralign with their discs by the time the
binary approaches coalescence. However this is not true for larger spins. In
particular, for $a>0.5$ we expect a sizeable fraction of the systems, of the
order of 30-40\% (or even more for extreme Kerr black holes), to  not reach the aligned configuration
after 10 Myrs. The result is still
present, for 50 Myrs, in which case obviously the black holes have more time
to become aligned.



The probability distributions computed with $\alpha=0.1$ are quite different from the others, because in this case we completely remove from the alignment process all the events with negative viscosity coefficient. As noted above, \citet{ogilvie99} predicts this behavior for small values of $\alpha$. We note however that even such a strong assumption does not affect significantly the distributions for large value of $a$ where the effects described above are definitely more important.

For all the other values of $\alpha$, there is no negative viscosity event: we note that our results are almost independent on $\alpha$ and that they are not very sensitive to the proxy chosen (10 or 50 Myrs), ensuring that our results are robust with respect to different choices of the merger timescale.

In Fig. \ref{fig:probs} we also plot with red lines (and using the same line styles as above) the corresponding probabilities in the case where we use the warp diffusion coefficient in the linear regime (Eq. 2) \citep{ogilvie99,LP10}  and we thus do not consider the effect of the non-linearity of the warp on the diffusion coefficients. The effect of the non-linearity is clearly very significant. In particular, it turns out that in the linear case the alignment probability has a much stronger dependence on $\alpha$ (essentially because in the non-linear case the range of values of $\alpha_2$ is much smaller than in the linear case) and alignment is indeed much faster, especially for low $\alpha$. 

We have also considered simpler cases where we keep the Eddington ratio fixed, in order to evaluate more clearly the dependance on this parameter. Clearly, since the alignment time is inversely proportional to $\dot{M}$, larger Eddington ratios (such as those assumed by BRM) imply a faster alignment. However, also in these cases, we find that alignment within 10Myrs occurs only for $a < 0.6$ and $a < 0.3$ for the two cases where $\dot{M}=0.1\dot{M}_{\rm Edd}$ and $\dot{M}=0.01\dot{M}_{\rm Edd}$, respectively. In order to have fast alignment for high spins the accretion rate is required to stay at the Eddington level for a prolonged period of time.

\section{Discussion and conclusions}
\label{sec:conclusions}

We have revisited the arguments suggesting that gas
disc are effective in bringing the spins of the two black holes in a
merging binary into alignment. In particular, we have improved on
previous estimates (BRM) by taking into full consideration how
the alignment timescale changes with the system parameters. In addition,
we have also included the reduction in the warp diffusion coefficient
when the misalignment angle becomes large \citep{ogilvie99}, an
effect previously neglected. 


Contrary to previous claims, if the black holes are rapidly spinning ({$a\gtrsim 0.5$), the
system is would not end up to be completely aligned, in up to 40\% of the cases,
at the time at which the two holes are brought together at distances of the order of 
0.01 pc (the current resolution limit for numerical simulations of the process). 

Our results are consistent, at least within orders of magnitude, with the previous investigation by \cite{perego09}. In particular, we find a similar dependence of the alignment timescale with respect to the spin of the black hole and the Eddington ratio. Contrary to \citet{perego09}, our estimates show that the alignment timescale is a strong function of the initial misalignment angle. This comes from having considered the full non linear warp propagation theory, instead of restricting to the small amplitude regime: this is a key element, because it allow us to compute a probability for the alignment process, based on the expected distribution of initial misalignments.

 Highly spinning black holes are thus more likely to maintain a significant misalignment, 
which causes a high recoil velocity. 
The current lack of observational evidence for strongly recoiling
black holes (but see the recent case of CID-42, \citealt{civano12})
 suggests that the average black hole spin is rather low, in line
with the predictions of the chaotic accretion picture \citep{king06}.
We note that for $a \lesssim 0.6$, as suggested above, the accretion
efficiency is $\epsilon \simeq 0.1$. This is in line with the values
suggested by the Soltan argument \citep{soltan82,yu02}
relating the average SMBH mass to the radiation background of the
universe (see also detailed discussion in \citealt{kph08}). We stress here that our conclusion regarding the magnitude of the black hole spin is related to the average properties of the BH population. Individual SMBH might well have large spins, and observations of broad iron lines (e.g., \citealt{brenneman06}) would naturally be biased in their favour. On the contrary, if recoiling black holes are found to be more common, such limitation on the magnitude of the black hole spin would not apply.

Our model is certainly very idealized and could be refined in several
ways. 
First of all, we have used a uniform logarithmic distribution of the Eddington value without considering in a detailed way how do tidal effects and gap
opening \citep{arty96} affect the accretion rate on the binary elements. Another interesting effect arises from the fact that the alignment timescale scales inversely with the accretion rate. In binaries with a large mass ratio, where accretion occurs preferentially onto the secondary, the primary might be much harder to align, with potentially interesting effects on gravitational wave emission.

A second important limitation of our work comes from the assumption
that the lengthscale over which the disc inclination varies is
comparable with the disc size. Also in this case, our assumptions go
in the direction of underestimating the alignment timescale, since if
the warp occurs over a short lengthscale even a relatively small
initial misalignment might be more difficult to realign. To take this
effect into account, one would need to explicitly solve for the disc
shape in a time--dependent calculation \citep[e.g.][]{LP06}.  As a consequence, we cannot quantify the exact degree of misalignment at the end of the merger, but can only perform a simple comparison between the two timescales. In order to quantify more precisely the level of misalignment, a detailed time-dependent calculation should be performed, along the lines of \citet{perego09} and \citet{dotti10}. Additionally, our choice of 10 and 50 Myrs as proxies for the shrinking timescale is clearly very simplified, and might in particular be a function of the system parameters, such as the accretion rate. For example, a large value of $t_{\rm align}$ can be due to a small value of $\dot{M}$ and should then be compared with a longer shrinking timescale, suitable for the expected gas-poor environment.

Third, we have considered only the effect of gas discs on the evolution of black hole spin. To follow the evolution of the spins after decoupling due to relativistic effects, a post-Newtonian approach is needed, taking into account the possible role of spin--orbit resonances \citep{schnittman04,berti12}.

Finally, we have
considered only the alignment process of one black hole with its own
disc. Aligment here is clearly a necessary condition for coalignment
of the two spins in a merging binary, Even in cases where we predict
alignment of individual spins and discs, the coalignment of the two
spins with the larger scale circumbinary disc can present additional severe difficulties if the black holes are rapidly spinning (see \citealt{nixon12} for a discussion).  All these effects act in the direction of reducing
the likelihood of alignment of both spins even further.
 
\section*{Acknowledgements}

We thank the referee, Tamara Bogdanovi{\'c}, and Chris Reynolds, Cole Miller and Massimo Dotti for their constructive criticism. We also thank Andrew King, Chris Nixon, Monica Colpi and Emanuele Berti for stimulating discussions and Gordon Ogilvie for
providing the code for computing the effective viscosities in a warped disc. DG was partially supported by NSF CAREER Grant No. PHY-1055103 and by the LIGO REU program at the California Institute of Technology.

\bibliography{lodato}

\begin{thebibliography}{36}
\expandafter\ifx\csname natexlab\endcsname\relax\def\natexlab#1{#1}\fi

\bibitem[{{Anderson} {et~al.}(2010){Anderson}, {Lehner}, {Megevand}, \&
  {Neilsen}}]{anderson10}
{Anderson}, M., {Lehner}, L., {Megevand}, M., \& {Neilsen}, D. 2010, Physical
  Review D, 81, 044004

\bibitem[Artymowicz \& Lubow(1996)]{arty96} Artymowicz, P., \& Lubow, S.~H.\ 1996, ApJL, 467, L77 

\bibitem[{{Bardeen}(1973)}]{bardeen73}
{Bardeen}, J.~M. 1973, in Black Holes (Les Astres Occlus), ed. C.~{Dewitt} \&
  B.~S. {Dewitt}, 215--239

\bibitem[{{Bardeen} \& {Petterson}(1975)}]{bardeen75}
{Bardeen}, J.~M. \& {Petterson}, J.~A. 1975, ApJ, 195, L65

\bibitem[{{Berti} {et~al.}(2012){Berti}, {Kesden}, \& {Sperhake}}]{berti12}
{Berti}, E., {Kesden}, M., \& {Sperhake}, U. 2012, ArXiv e-prints

\bibitem[{{Blecha} {et~al.}(2012){Blecha}, {Civano}, {Elvis}, \&
  {Loeb}}]{blecha12}
{Blecha}, L., {Civano}, F., {Elvis}, M., \& {Loeb}, A. 2012, ArXiv e-prints

\bibitem[{{Bogdanovi{\'c}} {et~al.}(2009){Bogdanovi{\'c}}, {Eracleous}, \&
  {Sigurdsson}}]{bogdanovic09}
{Bogdanovi{\'c}}, T., {Eracleous}, M., \& {Sigurdsson}, S. 2009, ApJ, 697, 288

\bibitem[{{Bogdanovi{\'c}} {et~al.}(2007){Bogdanovi{\'c}}, {Reynolds}, \&
  {Miller}}]{bogdanovic07}
{Bogdanovi{\'c}}, T., {Reynolds}, C.~S., \& {Miller}, M.~C. 2007, ApJ, 661,
  L147

\bibitem[Brenneman 
\& Reynolds(2006)]{brenneman06} Brenneman, L.~W., \& Reynolds, C.~S.\ 2006, ApJ, 652, 1028 

\bibitem[{{Campanelli} {et~al.}(2007{\natexlab{a}}){Campanelli}, {Lousto},
  {Zlochower}, \& {Merritt}}]{campanelli2007a}
{Campanelli}, M., {Lousto}, C., {Zlochower}, Y., \& {Merritt}, D.
  2007{\natexlab{a}}, ApJ, 659, L5

\bibitem[{{Campanelli} {et~al.}(2007{\natexlab{b}}){Campanelli}, {Lousto},
  {Zlochower}, \& {Merritt}}]{campanelli2007b}
{Campanelli}, M., {Lousto}, C.~O., {Zlochower}, Y., \& {Merritt}, D.
  2007{\natexlab{b}}, Physical Review Letters, 98, 231102

\bibitem[{{Civano} {et~al.}(2012){Civano}, {Elvis}, {Lanzuisi}, {Aldcroft},
  {Trichas}, {Bongiorno}, {Brusa}, {Blecha}, {Comastri}, {Loeb}, {Salvato},
  {Fruscione}, {Koekemoer}, {Komossa}, {Gilli}, {Mainieri}, {Piconcelli}, \&
  {Vignali}}]{civano12}
{Civano}, F. et al., C. 2012, ApJ, 752, 49

\bibitem[{{Corrales} {et~al.}(2010){Corrales}, {Haiman}, \&
  {MacFadyen}}]{corrales10}
{Corrales}, L.~R., {Haiman}, Z., \& {MacFadyen}, A. 2010, MNRAS, 404, 947

\bibitem[{{Dotti} {et~al.}(2009{\natexlab{a}}){Dotti}, {Montuori}, {Decarli},
  {Volonteri}, {Colpi}, \& {Haardt}}]{dotti09}
{Dotti}, M., {Montuori}, C., {Decarli}, R., {Volonteri}, M., {Colpi}, M., \&
  {Haardt}, F. 2009{\natexlab{a}}, MNRAS, 398, L73

\bibitem[{{Dotti} {et~al.}(2009{\natexlab{b}}){Dotti}, {Ruszkowski}, {Paredi},
  {Colpi}, {Volonteri}, \& {Haardt}}]{dotti09b}
{Dotti}, M., {Ruszkowski}, M., {Paredi}, L., {Colpi}, M., {Volonteri}, M., \&
  {Haardt}, F. 2009{\natexlab{b}}, MNRAS, 396, 1640

\bibitem[{{Dotti} {et~al.}(2010){Dotti}, {Volonteri}, {Perego}, {Colpi},
  {Ruszkowski}, \& {Haardt}}]{dotti10}
{Dotti}, M., {Volonteri}, M., {Perego}, A., {Colpi}, M., {Ruszkowski}, M., \&
  {Haardt}, F. 2010, MNRAS, 402, 682

\bibitem[{{Escala} {et~al.}(2005){Escala}, {Larson}, {Coppi}, \&
  {Mardones}}]{escala05}
{Escala}, A., {Larson}, R.~B., {Coppi}, P.~S., \& {Mardones}, D. 2005, ApJ,
  630, 152

\bibitem[{Fan {et~al.}(2004)}]{fan04}
Fan, X. {et~al.} 2004, AJ, 128, 515

\bibitem[{{Fan} {et~al.}(2006)}]{fan06}
{Fan}, X. {et~al.} 2006, AJ, 131, 1203

\bibitem[{{King} {et~al.}(2005){King}, {Lubow}, {Ogilvie}, \&
  {Pringle}}]{klop05}
{King}, A.~R., {Lubow}, S.~H., {Ogilvie}, G.~I., \& {Pringle}, J.~E. 2005,
  MNRAS, 363, 49

\bibitem[{{King} \& {Pringle}(2006)}]{king06}
{King}, A.~R. \& {Pringle}, J.~E. 2006, MNRAS, 373, L90

\bibitem[{{King} \& {Pringle}(2007)}]{king07}
---. 2007, MNRAS, 377, L25

\bibitem[{{King} {et~al.}(2008){King}, {Pringle}, \& {Hofmann}}]{kph08}
{King}, A.~R., {Pringle}, J.~E., \& {Hofmann}, J.~A. 2008, MNRAS, 385, 1621

\bibitem[{{Komossa}(2012)}]{komossa12}
{Komossa}, S. 2012, Advances in Astronomy, 2012

\bibitem[{{Komossa} {et~al.}(2008){Komossa}, {Zhou}, \& {Lu}}]{komossa08}
{Komossa}, S., {Zhou}, H., \& {Lu}, H. 2008, ApJ, 678, L81

\bibitem[Lodato 
\& Natarajan(2006)]{LN06} Lodato, G., \& Natarajan, P.\ 2006, MNRAS, 371, 1813 

\bibitem[{{Lodato} {et~al.}(2009){Lodato}, {Nayakshin}, {King}, \&
  {Pringle}}]{lodato09}
{Lodato}, G., {Nayakshin}, S., {King}, A.~R., \& {Pringle}, J.~E. 2009, MNRAS,
  398, 1392

\bibitem[{{Lodato} \& {Price}(2010)}]{LP10}
{Lodato}, G. \& {Price}, D.~J. 2010, MNRAS, 405, 1212

\bibitem[{{Lodato} \& {Pringle}(2006)}]{LP06}
{Lodato}, G. \& {Pringle}, J.~E. 2006, MNRAS, 368, 1196

\bibitem[{{Lousto} \& {Zlochower}(2011)}]{lousto11}
{Lousto}, C.~O. \& {Zlochower}, Y. 2011, Physical Review Letters, 107, 231102

\bibitem[{{Mortlock} {et~al.}(2011){Mortlock}, {Warren}, {Venemans}, {Patel},
  {Hewett}, {McMahon}, {Simpson}, {Theuns}, {Gonz{\'a}les-Solares}, {Adamson},
  {Dye}, {Hambly}, {Hirst}, {Irwin}, {Kuiper}, {Lawrence}, \&
  {R{\"o}ttgering}}]{mortlock11}
{Mortlock}, D.~J., et al. 2011,
  Nature, 474, 616

\bibitem[{{Natarajan} \& {Pringle}(1998)}]{natarajan98}
{Natarajan}, P. \& {Pringle}, J.~E. 1998, ApJ, 506, L97

\bibitem[{{Nixon}(2012)}]{n12}
{Nixon}, C.~J. 2012, MNRAS, 3054

\bibitem[{{Nixon} {et~al.}(2011{\natexlab{a}}){Nixon}, {Cossins}, {King}, \&
  {Pringle}}]{nixon11a}
{Nixon}, C.~J., {Cossins}, P.~J., {King}, A.~R., \& {Pringle}, J.~E. 2011{\natexlab{a}}, MNRAS, 412, 1591

\bibitem[{{Nixon} \& {King}(2012a)}]{NK12}
{Nixon}, C.~J., \& {King}, A.~R. 2012a, MNRAS, 421, 1201

\bibitem[{{Nixon} \& {King}(2012b)}]{nixon12}
{Nixon}, C.~J., \& {King}, A.~R. 2012b, submitted

\bibitem[{{Nixon} {et~al.}(2011{\natexlab{b}}){Nixon}, {King}, \&
  {Pringle}}]{nixon11b}
{Nixon}, C.~J., {King}, A.~R., \& {Pringle}, J.~E. 2011{\natexlab{b}}, MNRAS,
  L311+

\bibitem[{{Ogilvie}(1999)}]{ogilvie99}
{Ogilvie}, G.~I. 1999, MNRAS, 304, 557

\bibitem[{{Papaloizou} \& {Pringle}(1983)}]{pappringle83}
{Papaloizou}, J.~C.~B. \& {Pringle}, J.~E. 1983, MNRAS, 202, 1181

\bibitem[{{Perego} {et~al.}(2009){Perego}, {Dotti}, {Colpi}, \&
  {Volonteri}}]{perego09}
{Perego}, A., {Dotti}, M., {Colpi}, M., \& {Volonteri}, M. 2009, MNRAS, 399,
  2249

\bibitem[{{Rossi} {et~al.}(2010){Rossi}, {Lodato}, {Armitage}, {Pringle}, \&
  {King}}]{rossi10}
{Rossi}, E.~M., {Lodato}, G., {Armitage}, P.~J., {Pringle}, J.~E., \& {King},
  A.~R. 2010, MNRAS, 401, 2021

\bibitem[{{Scheuer} \& {Feiler}(1996)}]{sf96}
{Scheuer} P.~A.~G.,  {Feiler} R.,  1996, MNRAS, 282, 291

\bibitem[{{Schnittman}(2004)}]{schnittman04}
{Schnittman}, J.~D. 2004, Physical Review D, 70, 124020

\bibitem[{{Schnittman} \& {Krolik}(2008)}]{schnittman08}
{Schnittman}, J.~D. \& {Krolik}, J.~H. 2008, ApJ, 684, 835

\bibitem[{Shakura \& Sunyaev(1973)}]{shakura73}
Shakura, N.~I. \& Sunyaev, R.~A. 1973, A\&A, 24, 337

\bibitem[Soltan(1982)]{soltan82} Soltan, A.\ 1982, MNRAS, 200, 
115 

\bibitem[{{Volonteri} \& {Rees}(2005)}]{VR2005}
{Volonteri}, M., \& {Rees}, M.~J. 2005, ApJ, 633, 624

\bibitem[Yu \& Tremaine(2002)]{yu02} Yu, Q., \& Tremaine, S.\ 2002, MNRAS, 335, 965 

\end{thebibliography}
\label{lastpage}
\end{document}